\documentstyle[eqsecnum,aps,epsfig,eepic]{revtex}
\newcommand{\nc}{\newcommand}
\nc{\ba}{\begin{array}}
\nc{\ea}{\end{array}}
\nc{\bml}{\begin{mathletters}}
\nc{\eml}{\end{mathletters}}
\nc{\bfi}{\begin{figure}}
\nc{\efi}{\end{figure}}
\nc{\nn}{\nonumber}
\nc{\pp}{{\prime\prime}}
\nc{\lav}{\left<\!\left<}
\nc{\rav}{\right>\!\right>}
\nc{\hf}{\case{1}{2}}
\nc{\ptl}{\partial}
\nc{\eg}{, {\em e.~g.},~}
\nc{\egb}{$e.\ g.$~}
\nc{\ie}{, {\em i.~e.},~}
\nc{\rhs}{r.\ h.\ s.\ }
\nc{\lhs}{l.\ h.\ s.\ }
\nc{\wrt}{w.\ r.\ t.\ }
\nc{\lamb}{\mbox{\boldmath $\lambda$}}
\nc{\bsigma}{\mbox{\boldmath $\sigma$}}
\nc{\mat}{\sf}
\begin{document}
\draft 
 
\title{~ \\ ~ \\ ~\\ ~ \\ Beyond storage capacity in a single model
neuron: Continuous replica symmetry breaking }
 
\author{G. Gy\"orgyi } 
 
\address{Institute for Theoretical Physics, E\"otv\"os University \\
1518 Budapest, Pf.\ 32, Hungary, e-mail: gyorgyi@glu.elte.hu} 
 
\author{P. Reimann} \address{Theoretische Physik, Universit\"at
Augsburg, 86135 Augsburg, Germany \\
e-mail: reimann@physik.uni-augsburg.de}
   
\date{\today} \maketitle \widetext
\begin{abstract} 
 
A single McCulloch-Pitts neuron, that is, the simple perceptron is
studied, with focus on the region beyond storage capacity.  It is
shown that Parisi's hierarchical ansatz for the overlap matrix of the
synaptic couplings with so called continuous replica symmetry breaking
is a solution, and as we propose it is the exact one, to the
equilibrium problem.  We describe some of the most salient features of
the theory and give results about the low temperature region.  In
particular, the basics of the Parisi technique and the way to
calculate thermodynamical expectation values is explained.  We have
numerically extremized the replica free energy functional for some
parameter settings, and thus obtained the order parameter function\ie
the probability distribution of overlaps.  That enabled us to evaluate
the probability density of the local stability parameter.  We also
performed a simulation and found a local stability density closer to
the theoretical curve than previous numerical results were.

\end{abstract}  
 
\vspace{1cm} Keywords: {\em neural networks, perceptron, storage
capacity, replica method} \vspace{1cm}
 
 
\section{Introduction}  
\label{sec_intro} 
 
In his seminal paper Hopfield \cite{hop82} reformulated Little's model
\cite{lit74} of auto-associative memory in terms of an energy
function.  By this act, the field of the statistical mechanics of
neural networks was plowed and sown in, and proved itself since then
remarkably fertile.  The network is an interconnected set of
McCulloch-Pitts neurons \cite{mcc43}, the latter being perhaps the
biologically least realistic model of a nerve cell.  In its simplest
version the model neuron can be in one of only two states, ``firing''
or ``quiescent'', it is nonlinear, and admits a large number of
connections from other units.  Despite the oversimplification in its
node cells, the network can exhibit complex behavior and functions as
a reasonable model of content addressable memory.  From the practical
viewpoint, however, artificial neural networks are generically not
superior to other methods in the task of storage and associative
retrieval \cite{eg88,hkp91}.
 
An important ingredient of this type of models of neural networks is
what is called in statistical mechanics quenched disorder.  For
example, in the Little-Hopfield model the synaptic coupling strengths,
made up of random numbers by the Hebb rule \cite{sgrev87,hkp91}, can
be held fixed for the duration of the neural dynamical process.  This
is the basis for the analogy between spin glass models, the
archetypical example of a many-body problem with quenched disorder,
and neural networks \cite{sgrev87,sgrev91}.  Methods borrowed from the
theory of spin glasses, in particular from the infinite range
interaction Sherrington-Kirkpatrick model, yielded a harvest of
results on a variety of neural model systems, as well as on other
problems whose motivation came from outside of physics but could be
formulated as disordered statistical mechanical systems
\cite{hkp91,sgrev87,sgrev91}.
 
It is safe to say that the technique inherited from spin glass theory
and the most widely used in the statistical mechanical approach to
neural networks in equilibrium is the replica method \cite{sgrev87}.
It was first applied thoroughly for infinite range interaction spin
glasses, thus it is especially suited for networks where neurons have
a large number of connections.  In its simplest version, with so
called replica symmetry, it can be straightforwardly adopted to many a
neural problem \cite{hkp91}.
  
Another family of artificial neural networks consists of layered
feed-forward networks \cite{ros62}, which accept a number of inputs
and for a given set of couplings produce the output as a function of
the inputs.  Such networks were introduced for the purpose of
generalization\ie rule extraction, from examples of input-output
pairs.  Then the {\em a priori} unknown target rule is to be
reconstructed by adaptive changes in the couplings, called training.
Since the introduction of the backpropagation algorithm \cite{rum86a}
for that purpose, feed-forward networks found wide usage
\cite{hkp91,wat93,ama98}.
  
The statistical physics of neural modeling gained an impetus of
lasting effect from the work by Gardner and Derrida \cite{eg88,gd88}
on the storage problem of a single neuron, called also simple
perceptron in that context.  This is the simplest version \cite{min69}
of feedforward networks.  While in the Little-Hopfield statistical
mechanical system the quenched variables are the couplings and a
microstate is a configuration of neural states, the roles in
feed-forward networks are reversed.  Now the space of synaptic
couplings is considered as the configuration space within Boltzmannian
thermodynamics and the examples appear as quenched parameters.  The
error for one example measures the difference between the actual
output of the network and the required output.  The errors on all
training examples add up to form the Hamiltonian\ie the cost function,
of the statistical mechanical model.  Within the canonical statistical
mechanical approach the temperature plays its usual roles. On the one
hand, it is the Lagrange multiplier associated with a preset value of
the error, on the other hand, it is the amplitude of noise if a
gradient-descent-like dynamics of the couplings is used so as to reach
optimal configuration.  The thermodynamical limit is achieved by
admitting a large number of adjustable couplings, but for that it is
not necessary to have many neurons.  The approach of Gardner and
Derrida was successful in what is called equilibrium learning, when a
Hamiltonian can be associated with the problem.  However, statistical
physical methods are proving themselves useful also in studying
off-equilibrium learning algorithms \cite{bc99}.
 
A central quantity of a feed-forward network is its storage
capacity\ie how many random input-output examples the network can
reproduce without error.  In terms of the statistical mechanical
approach this is in its original formulation a zero temperature
problem.  The subspace of couplings that reproduce a given set of
patterns is called version space, its volume, related to the ground
state ($T=0$) entropy, vanishes beyond capacity.  Since the
statistical mechanical solution of the region below capacity of a
single neuron by Gardner and Derrida \cite{eg88,gd88} a number of
results have been obtained about storage properties of feed-forward
networks, see for example Refs.\ \cite{ama98,mal97,sch95,wes98,win97}.
Nevertheless, if the task is to minimize the number of incorrectly
stored examples, beyond capacity the problem has not been solved.
Technically this is because below capacity for completely random
examples replica symmetry holds, while beyond it no finite replica
symmetry breaking scheme yields thermodynamically stable solution
\cite{ws96}. 
 
In the present paper we reconsider the problem of storage of random
patterns, technically generalize Parisi's solution of the
Sherrington-Kirkpatrick model, and obtain beyond capacity a phase
reminiscent to the frustrated ground state of the
Sherrington-Kirkpatrick model.  That phase continues for $T>0$ into
the analog of the low temperature, or Parisi, phase of the spin glass.
 
Our work is motivated not only by the problem of storing random
patterns.  Generalization has also been successfully analyzed by
statistical mechanical methods, see Ref.\ \cite{wat93,ok96} for
reviews.  The storage problem below capacity is analogous to
equilibrium learning of a learnable task, where the network is
compatible with all possible examples, there is no frustration in
either systems.  For instance, equilibrium generalization properties
of the perceptron when the examples are generated by another one, can
be understood for arbitrary number of examples within replica symmetry
\cite{gt90}.  However, a given feed-forward network may be unable to
reproduce a complicated target function on all possible examples
\cite{wat93}.  If the target is unlearnable then the network is
presumed to get into a frustrated phase if a sufficiently large number
of examples are used.  Thus the properties of a network beyond
capacity are of foremost interest from the viewpoint of rule
extraction as well.
  
The single neuron may be considered as the hydrogen atom of neural
problems and studied for its own interest.  It is the unit of the
Little-Hopfield network, where the symmetry in the couplings has been
given up and the couplings of different neurons are considered as
independent.  As to a feed-forward network, even if the whole network
operates without error, its units may still be strained beyond their
individual capacities \cite{mal97}.  Thus the description of a single
neuron beyond its storage capacity is of importance also from the
viewpoint of networked neurons. Furthermore, a close analogy exists
between the behavior of model neurons beyond capacity and the glassy,
frustrated, phase of disordered spin systems
\cite{sgrev87,sgrev91,mez93,et93,ws96,our-paper}. Therefore, the
understanding of the way a single neuron works may have ramifications
beyond the field of artificial neural networks.
 
We firstly review in Sec.\ \ref{sec_storage} the statistical mechanics
of storage, recall the basic thermodynamical quantities and the
formula for the replica free energy of a single neuron.  In Sec.\
\ref{sec-ps} we summarize the main ingredients of Parisi's approach
and obtain the free energy functional that we propose describes the
equilibrium problem.  This way some background is given to our
previous communication \cite{our-paper}, wherein we identified a spin
glass phase of Parisi type in the high temperature limit.  The recipe
for the calculation of expectation values by means of Green functions
is explained in Sec.\ \ref{sec-st-exp}, producing among other the
formulas for the local stability distribution and the stationarity
conditions.  Sec.\ \ref{lt} is devoted to the scaling for low
temperatures, enabling us to put the extremization of the free energy
functional on a computer, and in the end a simulation is discussed. 
 
\section{The  storage problem and its replica free energy}  
\label{sec_storage} 
 
The model neuron, or perceptron, under consideration is
\cite{mcc43,ros62} \bml
\label{defneu}
\begin{eqnarray}
   \label{defneu1}
   \xi & = & \text{sign}(h),  \\
   \label{defneu2}
   h & =  & N^{-1/2} \sum\nolimits_{k=1}^{N} J_k S_k,  
\end{eqnarray}
\eml where $\bf J$ is the synaptic coupling vector, $\bf S$ the input
and $\xi$ the output.  Patterns are given as input-output data,
\begin{equation} \label{patterns}
\{ {\bf S}^\mu ,\xi ^\mu\}_{\mu =1}^{M}.
\end{equation}
In the simplest setup the $S_k^\mu$-s are independently drawn from any
distribution with unit variance and zero average, and $\xi^\mu = \pm
1$ with equal probability.  We introduce the local stability parameter
as
\begin{eqnarray}
   \label{def-delta}
\Delta^\mu = h^\mu \xi^\mu,
\end{eqnarray} 
where $h^\mu$ is given by (\ref{defneu2}) with $S_k^\mu$.  If the
neuron generates ${\xi^\mu}$ in response to ${\bf S}^\mu$\ie
$\Delta^\mu >0$, then we say that the $\mu$-th pattern is stored.  If
$\Delta^\mu$ is a large positive number then high stability of storage
against changes in either the couplings or the inputs can be assumed.
Large stability is associated with large basin of attraction in memory
networks \cite{hkp91}.  Given the ensemble of patterns, the local
stability parameter obeys some distribution $\rho(\Delta)$
\cite{gg91}.  If the number of patterns $M$ is of order $N$ then it is
useful to introduce the relative number of examples
\begin{equation} \label{def-alpha}
\alpha = M/N,
\end{equation}
called also load parameter.  Since an overall positive factor of $\bf J$ does
not change the output, we set the norm of $\bf J$ to $\sqrt{N}$,
expressed by the prior distribution
\begin{equation}\
  \label{spher-prior}
  w({\bf J}) =   C_N \ \delta\left(N-\left|{\bf J}\right|^2\right).
\end{equation}
This is called spherical constraint.  The factor $C_N$ normalizes
$w({\bf J})$ to unity, it has no thermodynamical significance besides
setting the zero point of the entropy scale.  Given the distribution
of patterns and the length of $\bf J$ it can be easily seen that the
normalization in (\ref{defneu2}) results in $h$ values of typically
$O(1)$.
 
Storage with minimal error can be formulated as an optimization task
by our introducing an error measure.  If we treat all patterns in the
same way we obtain what is called the equilibrium problem.  The
associated Hamiltonian, or cost function, is
\begin{equation}
  \label{hamiltonian} {\cal H} = \sum_{\mu =1}^M V(\Delta^\mu),
\end{equation}
where the potential $V(\Delta^\mu)$ gives the error on a single
pattern ${\bf S}^\mu ,\xi ^\mu$.  Obviously, $V(y)=0$ for $y>0$ in the
original storage problem.  One can also impose a bound $\kappa$ for
the local stability\ie $V(y)$ is set to be zero for $y>\kappa$, if
$\kappa>0$ this is obviously stricter than the original storage
criterion.  Generically, $V(y)$ should be monotonically decreasing for
$y<\kappa$.  In this paper specific results will be presented on the
error counting, or Gardner-Derrida \cite{eg88,gd88}, potential
\begin{equation}
  \label{gen-v}
  V(y)=\theta (\kappa -y),
\end{equation}
where $\theta(y)$ is the Heaviside function.  Given the potential
(\ref{gen-v}) the aim is to minimize the number of patterns whose
stability is below the bound $\kappa$.  In the theoretical framework
we shall keep the general form $V(y)$.

The partition function of the optimization task is 
\begin{equation}
  \label{partfunc}
   Z = \int d^N\!  J\ w({\bf J}) \ \exp \left(-\beta \sum_{\mu =1}^M
V(\Delta^\mu)\right), 
\end{equation}
with $\beta=1/T$.  Quenched average,
$\langle\dots\rangle_{\text{qu}}$, is defined as the mean over the
patterns.  We deal with the idealized equilibrium of the system, when
for large $N$ the free energy, the energy, and the entropy are assumed
to approach their quenched average.  This property of self-averaging
was proved rigorously only in special cases, see for example
\cite{aiz87}, but it is the widely used basis in studies of the
equilibrium thermodynamics of disordered systems \cite{sgrev87}.

The replica method \cite{sgrev87,sgrev91} consists in writing the mean
free energy per coupling as
\begin{equation} 
  \label{fe} f = - \lim_{N\to\infty}\frac{\left<\ln Z\right>_{\text
  {qu}}}{N\beta} =\lim_{N\to\infty} \lim_{n\rightarrow 0}
  \frac{1-\left<Z^n\right>_{\text{qu}}}{nN\beta}.
\end{equation} 
Denoting the thermal average with the Boltzmann weight in
(\ref{partfunc}) by $\langle\dots\rangle_{\text{th}}$ the mean error
per pattern can be written as
\begin{equation}  
  \label{ener-def} \varepsilon = \left< \left< V(\Delta)
  \right>_{\text{th}} \right>_{\text{qu}}=
  \frac{1}{\alpha}\frac{\partial\beta f}{\partial\beta}.
\end{equation} 
The entropy is
\begin{equation} 
  \label{entr} 
  s = \beta(\alpha\varepsilon - f).  
\end{equation} 
For $T=0$ and $\varepsilon =0$ the volume of version space\ie the
space of couplings that perfectly reproduce the examples is obtained
as $\Omega=\exp(Ns)$.  In general, $Ns$ has the usual meaning of the
logarithm of the volume with given error $\varepsilon$.

Introducing the overlap matrix $\sf Q$ of synaptic vectors
\begin{equation}
  \label{syn-overlap}
  \left[{\sf Q}\right]_{ab} \equiv q_{ab}
  = \frac{1}{N} \sum_{k=1}^N J_{ak}J_{bk},
\end{equation}
where the replica indices $a,b$ go from $1$ to $n$, we can express the
free energy as the result of the minimization of the replica free energy
\cite{gg91,mez93,our-paper} \bml
\label{fe-spher1}
  \begin{eqnarray}
  \label{fe-spher1a}
  f & = & \lim_{n\rightarrow 0} \frac{1}{n}~ {\min_{\sf Q}}
  \ f({\sf  Q}) \\
  \label{fe-spher1b} 
  f({\sf Q}) & = & f_s({\sf Q}) + \alpha\, f_e({\sf Q}) ,\\
  \label{fe-spher1c}
  f_s({\sf Q}) & = & - (2\beta)^{-1} \ln\text{det}{\bf\sf Q}, \\
  \label{fe-spher1d}
  f_e({\sf Q}) & = & - \frac{1}{\beta}\ln \int \frac{d^n\! x \ d^n\!
    y}{(2\pi)^n} \exp\left( -\beta
    \sum\nolimits_{a=1}^n V(y_a) + i{\bf xy} - \case{1}{2} 
  {\bf x} {\sf Q}{\bf x}\right).
\end{eqnarray}
\eml The subscripts $s$ and $e$ stand for entropic and energy-like
terms.  The entropic contribution (\ref{fe-spher1c}) arises because of
the spherical constraint and the definition (\ref{syn-overlap}),  it
is indeed independent of the potential, while (\ref{fe-spher1d})
depends on it.

A central role is played by the probability density for the local
stabilities \cite{gg91,mez93}
\begin{equation}
  \label{def-dist-stab}
  \rho(\Delta) = \left< \left< \delta \left( \Delta -h^1\xi^1\right)
      \right>_{\text{th}}\right>_{\text{qu}},
\end{equation}
where (\ref{defneu2}) is understood.  Due to the symmetry of the
Hamiltonian with respect to the permutation of patterns we could
choose $\mu=1$ for convenience.  It will turn out to be useful to
interpret the integrand in (\ref{fe-spher1d}) as an effective
Boltzmann weight and denote the average over this measure as
\begin{equation}
\label{av}
\lav\dots\rav,
\end{equation}
where the $n\to 0$ limit is implied.  A straightforward replica
calculation shows (see Ref.\ \cite{our-paper2} for a pedagogic
presentation) that the local stability distribution can be rewritten
as
\begin{equation}
  \label{repl-dist-stab}
   \rho(\Delta) = \lav \delta \left( \Delta - y_1 \right)
   \rav .
\end{equation}
Here the subscript could be any replica index, for convenience we
chose $1$.  Comparison with (\ref{def-dist-stab}) allows an intuitive
interpretation for the replica average $\lav \dots \rav$, namely, this
corresponds to the combined average over thermal and quenched
fluctuations.  From (\ref{def-dist-stab},\ref{repl-dist-stab}) it
follows that the combined thermal and quenched average in
(\ref{ener-def}) boils down to
\begin{equation} 
  \label{ener-dist} \varepsilon = \lav V \left( y_1 \right)
  \rav  = \int dy \ \rho(y)\, V(y).
\end{equation}

The free energy (\ref{fe-spher1}) was calculated within the replica
symmetric ansatz for the error counting potential (\ref{gen-v}) and
the capacity\ie the maximal $\alpha$ with $\varepsilon=0$ at $T=0$ was
determined in \cite{eg88,gd88}.  It has been shown that beyond
capacity the replica symmetric solution is thermodynamically unstable
\cite{bou94}.  One \cite{mez93,et93} and two \cite{ws96} step replica
symmetry breaking solutions were presented, while Ref.\ \cite{ws96}
proved that no finite step symmetry breaking ansatz can possibly be
thermodynamically stable.  We presented in \cite{our-paper} a
variational free energy functional without derivation that
incorporated continuous replica symmetry breaking, but gave concrete
results only in the high temperature, large $\alpha$ limit.  In what
follows we provide some background to the general theory and in the
end properties of the ground state ($T=0$) beyond capacity will also
be described.

\section{The Parisi scheme}
\label{sec-ps}

In this section we show how to evaluate the replica free energy
$f_e({\sf Q})$ of (\ref{fe-spher1d}) within Parisi's ansatz. The $R$
step replica symmetry breaking form is \cite{par80a}
\begin{equation}
  \label{pa}
  {\sf Q} = \sum_{r=0}^{R+1} \left( q_r-q_{r-1}\right)  {\sf U}_{m_r}
\otimes  {\sf I}_{n/m_r},
\end{equation}
where $k$ subscripts $k$-dimensional matrices, ${\sf I}_{k}$ is the
identity operator, all elements of ${\sf U}_{k}$ equal $1$, $\otimes$
marks the direct product, and 
\bml\label{ieq}
\begin{eqnarray}
  \label{ieq0} q_{-1} & = & 0 \leq q_0 \leq q_1 \dots \leq q_R \leq
  q_{R+1} =1,\\ \label{ieq2} m_{R+1} & = & 1 \leq m_R
  \leq m_{R-1} \dots \leq m_1 \leq m_0 = n,
\end{eqnarray}
\eml where the integer $m_r$ is a divisor of $m_{r-1}$.  The $n\to 0$
limit can be performed smoothly if instead of $m_r$ we use
$x_r=(n-m_r)/(n-1)$ for parametrization.  Thus for arbitrary $n>0$ we
have the ordering
\begin{equation}
  \label{ieq2b}
   x_{R+1} =  1 \geq x_R \geq x_{R-1} \dots \geq x_1 \geq x_0 = 0.
\end{equation}
We consider the $x_r$-s fixed along the $n\to 0$ limiting process,
whence follows the formal $n$-dependence of the $m_r(n)$-s, and for
$n=0$ we get $x_r=m_r(0)$.  The inspection of the first few $R=0,1,2$
cases \cite{eg88,mez93,et93,ws96} allows, in the spirit of Parisi's
\cite{par80a}, the generalization of the energy term
(\ref{fe-spher1d}) in the replica free energy to arbitrary $R$ as
\begin{eqnarray}
 \label{phirrsb} f_e({\bf q},{\bf x}) & = & \lim_{n\to 0} \frac{1 }{n}
 f_e({\sf Q}) \nn \\ &=& -\frac{1}{\beta x_1}\int Dz_0 \ln \int Dz_1
 \nn \\ && \times \left[ \int Dz_2 \dots \left[ \int
 Dz_{R+1}\exp\left\{ -\beta V\left(\sum_{r=0}^{R+1} z_r
 \sqrt{q_r-q_{r-1}} \right)\right\} \right] ^\frac{x_R}{x_{R+1}} \dots
 \right]^\frac{x_{1}}{x_{2}},
\end{eqnarray}
where 
\begin{equation}
 \label{dz}
Dz = \frac{dz~e^{-\hf z^2}}{\sqrt{2\pi}}.
\end{equation}
This is the analog of Parisi's formula for the Sherrington-Kirkpatrick
model, Eq.\ (11) in \cite{par80a}; a comprehensive derivation will be
presented elsewhere \cite{our-paper2}.  The energy term
(\ref{fe-spher1d}) has become a function of the parameters in
(\ref{ieq0},\ref{ieq2b}).  The evaluation of (\ref{phirrsb}) can be
done by iteration, \bml \label{rec-psi}
\begin{eqnarray} 
  \label{rec-psi1} \psi_{r-1}(y) & = & \int Dz~
  \psi_r\left(y+z\sqrt{q_r-q_{r-1}}\right) ^ \frac{x_r}{x_{r+1}}, \\
  \label{rec-psi2}  
 \psi_{R+1}(y) & = & e^{-\beta V(y)},
\end{eqnarray} 
\eml 
where $x_{R+2}=1$ is understood.  Then the sought free energy term is
obtained as
\begin{equation}
  \label{ferecg} f_e({\bf q},{\bf x}) = -\frac{1}{\beta x_1} \int Dz~
  \ln \psi_0\left(z\sqrt{q_0}\right).
\end{equation}
where ${\bf q} = (q_0,\dots,q_R)$ and ${\bf x} = (x_1,\dots,x_R)$.

The above iteration can be redressed as a partial differential
equation.  Parisi's order parameter function $x(q)$ is a concatenation
of the $\bf q$ and $\bf x$ as
\begin{equation}
  \label{xq}
 x(q)  = \sum_{r=0}^R (x_{r+1}-x_r)\, \theta(q-q_r),
\end{equation}
where $x_{-1}=0$.  Next we introduce $\psi(q,y)$ such that at the
discontinuities
\bml
\begin{eqnarray}
  \label{psiplus} \psi(q_r^{+0},y) & = & \psi_r(y),\\ \label{psiminus}
 \psi(q_r^{-0},y) & = &
 \psi(q_r^{+0},y)^\frac{x(q_r^{-0})}{x(q_r^{+0})},
\end{eqnarray}
\eml that is, $\psi(q,y)^{1/x(q)}$ is continuous in $q$.  Furthermore,
along the plateau in the open interval $(q_{r-1},q_{r})$
\begin{equation}
\label{psi-cont}
  \psi(q,y) = \int Dz~ \psi\left(q_r^{-0},y+z\sqrt{q_r-q}\right).
\end{equation}
It is easy to show that the field so defined satisfies the partial
differential equation (PDE) 
\bml \label{psi-pde} 
\begin{eqnarray}
\label{psi-pde1} \ptl _q \psi(q,y) & = & -\hf \ptl _y ^2 \psi(q,y) +
\frac{\dot x(q) }{x(q)} \psi(q,y) \ln \psi(q,y), \\ \label{psi-pde2}
\psi(1,y) & = & e^{-\beta V(y)},
\end{eqnarray}
\eml which evolves from $q=1$ to $q=0$.  Indeed, along the plateaus
$\dot x(q)=0$ when only the first term on the \rhs of (\ref{psi-pde1})
remains, thus producing (\ref{psi-cont}).  Near jumps of $x(q)$ the
second term dominates, and at a fixed $y$ the resulting ordinary
differential equation in the variable $q$ is separable.  Hence it
follows that $\psi(q,y)^{1/x(q)}$ is continuous in $q$, thus
(\ref{psiminus}) is recovered.  An equivalent field can be defined by
\begin{equation}
  \label{psi-to-phi}
   f(q,y) = - \frac{\ln\psi(q,y)}{\beta x(q)},
\end{equation}
satisfying for a continuous potential $V(y)$ the PDE 
\bml \label{ppde}
\begin{eqnarray}
  \label{ppde1} \ptl _q f(q,y) & = & -\hf \ptl _y ^2 f(q,y) + \hf
   \beta x(q) \left(\ptl _y f(q,y)\right)^2 ,\\ \label{ppde2} f(1,y) &
   = & V(y).
\end{eqnarray} 
\eml The fact that $f$ denotes the free energy (\ref{fe-spher1a}) as
well as the field $f(q,y)$ should not cause misunderstandings.
Equation (\ref{ppde1}) with initial condition $\ln 2 \cosh \beta y$
has been discovered by Parisi \cite{par80a} while studying the
Sherrington-Kirkpatrick model.

If the potential $V(y)$ is not continuous, the PDE (\ref{ppde}) holds
only from any $q^\ast<1$ onward where $\ln\psi(q^\ast,y)$ is
continuous in $y$.  In the generic case such is $q_R$, so the
evolution along the first plateau from $1$ to $q_R$, where $x(q)\equiv
1$, is to be done explicitly as
\begin{equation}
  \label{rec-phi2} f(q_R,y) = - \frac{\ln \psi(q_R,y)}{\beta x_{R+1}}
  = - \beta^{-1} \ln \int Dz~ e^{- \beta V\left(y+z
  \sqrt{1-q_{R}}\right)},
\end{equation} 
and for $q<q_R$ the PDE (\ref{ppde1}) with the initial condition
(\ref{rec-phi2}) can be used.  Alternatively, one can introduce the
field $m(q,y)$ as 
\begin{equation}
\label{m}
  \psi(q,y) m(q,y)=-\frac{\ptl _y \psi(q,y)}{\beta x(q)},
\end{equation} 
which equals $\ptl _yf(q,y)$ when continuity in $y$ holds.  Then the
PDE (\ref{ppde1}) should be replaced by
\begin{equation}
\label{ppde-discontinuity}
 \ptl _q f(q,y)  =  -\hf \ptl _y m(q,y) + \hf \beta x m^2(q,y),
\end{equation} 
while the initial condition (\ref{ppde2}) can be kept.

Finally we obtain the sought free energy term (\ref{fe-spher1d}) as a
functional of the order parameter function
\begin{equation} 
  \label{fe-from-pde} f_e[x(q)] = \lim_{n\to 0} \frac{1}{n} f_e({\sf
  Q}) = f(0,0).
\end{equation} 
  
It should be emphasized that the above PDE-s do not require infinite
refining of the partition by $q_r$-s of the interval $(0,1)$.  They
are valid for discrete as well as continuous replica symmetry breaking
schemes\ie they admit $x(q)$ with steps and plateaus, as well as
strictly monotonically increasing continuous segments.
 
The entropic term (\ref{fe-spher1c}) can also be cast in the form of
the energy term (\ref{fe-spher1d}) with the substitution
\begin{equation}
  \label{spher-assoc}
  e^{-\beta V(y)} = \sqrt{2\pi}\, \delta(y).
\end{equation}
If we conceive the Dirac delta as a Gaussian with small variance, the
initial condition (\ref{ppde2}) becomes a quadratic function and the
PDE (\ref{ppde1}) can be solved analytically.  The analogue of
(\ref{fe-from-pde}) for the entropic term, after going with the
variance to zero, can be cast into
\begin{equation}
  \label{fes-spher-parisi}
  f_s[x(q)] =  \lim_{n\to 0} \frac{1}{n}  f_s({\sf Q}) 
  = - \frac{1}{2\beta} \int_0^1  
  dq\, \left[ \frac{1}{D(q)} - \frac{1}{1-q}\right],
\end{equation} 
where 
\begin{equation}
  \label{pa-cont-spectrum}
  D(q) = \int_q^1 d\bar{q} ~ x(\bar{q}).
\end{equation}

The free energy of the neuron is then obtained as
\begin{equation}
f =  {\ba{c}\mbox{\footnotesize ~} \\
  \label{fe-spher-var-ext} \mbox{max}\\ \mbox{\footnotesize{\em
  x(q)}}\ea} f\left[ x(q) \right],
\end{equation}
where the free energy functional is 
\begin{equation}
  \label{fe-functional-spher-def-ext} f[x(q)] = f_s[x(q)] + \alpha
  f_e[x(q)], 
\end{equation}
with $f_s$ and $f_e$ defined in Eqs.\
(\ref{fes-spher-parisi},\ref{fe-from-pde}).  It is due to the $n\to 0$
limit that the maximization in (\ref{fe-spher-var-ext}) replaces the
minimization by the matrix elements of $\sf Q$ in (\ref{fe-spher1a}),
see\eg Ref.\ \cite{sgrev87}.

\section{Linear response theory, stationarity conditions, and
 expectation values}
\label{sec-st-exp} 
 
The least obvious part of the extremization condition
(\ref{fe-spher-var-ext}) is the variation of $f(0,0)$ by $x(q)$.  This
can be calculated from linear response theory for the PDE
(\ref{ppde1}).  Moreover, linear response theory yields a technique
to calculate replica averages as introduced in (\ref{av}), essential
for the evaluation of physical quantities.

The Green function for the PDE (\ref{ppde1}) can be introduced
formally as
\begin{equation} 
 \label{gf-X} {\cal G}(q_1,y_1;q_2,y_2) = \frac{\delta
 f(q_1,y_1)}{\delta f(q_2,y_2)},
\end{equation}
whence ${\cal G}(q_1,y_1;q_2,y_2) = 0$ for $q_1>q_2$.  The Green
function for the Parisi solution of the Sherrington-Kirkpatrick model
has been studied in Refs.\ \cite{dl83,sd84}.  In the fore and hind
variable pairs the Green function ${\cal G}(q_1,y_1;q_2,y_2)$
satisfies the respective PDE-s 
\bml
\label{gf-pde}
\begin{eqnarray} 
  \label{gf-pde-forw} \ptl _{q_{1}} {\cal G} = -\hf\ptl _{y_{1}}^2
   {\cal G} + \beta x(q_1)m(q_1,y_1) \ptl _{y_1} {\cal
   G} - \delta(q_1-q_2)\delta(y_1-y_2), \\ \label{gf-pde-backw} \ptl
   _{q_{2}} {\cal G} = \hf\ptl _{y_{2}}^2 {\cal G} + \beta x(q_2) \ptl
   _{y_2} \left[ m(q_2,y_2) {\cal G}\right] +
   \delta(q_1-q_2)\delta(y_1-y_2),
\end{eqnarray} 
\eml where $m$ is given in (\ref{m}).  The first equation without the
Dirac delta excitation is the linearization of the PDE (\ref{ppde1}).
The minus sign of the Dirac deltas follows from the fact that
({\ref{gf-pde-forw}) evolves towards decreasing ``time'' $q$.  The
homogeneous part of (\ref{gf-pde-backw}) is obtained from the
requirement that
\begin{equation} 
 \label{gfs} {\cal G}(q_1,y_1;q_3,y_3) = \int dy_2\, {\cal
 G}(q_1,y_1;q_2,y_2)\,  {\cal G}(q_2,y_2;q_3,y_3)
\end{equation} 
does not depend on $q_2$, and the plus sign of the inhomogeneous term
is due to the fact that evolution goes towards increasing $q$.  The
homogeneous parts of two PDE-s (\ref{gf-pde-backw}) and
(\ref{gf-pde-backw}) are called adjoint to each other.

For the sake of simplicity we gave formula (\ref{gf-X}) for the case
of continuous potential $V(y)$.  If the potential is discontinuous
then the definition (\ref{gf-X}) should and can be appropriately
modified when a $q$-argument is near $1$, but the PDE-s (\ref{gf-pde})
for the Green functions hold as they are.
 
The significance of the Green function is in that it helps to solve
the linear PDE with the source term $h(q,y)$
\begin{equation}
  \label{theta-with-source-pde} \ptl _q \vartheta(q,y) = -\hf\ptl
   _y^2\vartheta(q,y) + \beta x(q) m(q,y) \ptl _y\vartheta(q,y) +
   h(q,y),
\end{equation}
as
\begin{equation}
 \label{theta-with-source-solve}
 \vartheta(q,y) = \int dy_1 {\cal G}_{\varphi}(q,y;1,y_1)
 \vartheta(1,y_1) - \int_q^1 dq_1 \int dy_1 {\cal
 G}_{\varphi}(q,y;q_1,y_1) h(q_1,y_1).
\end{equation}
A prominent role will be played by
\begin{equation} 
 \label{p} P(q,y) = {\cal G}(0,0;q,y),
\end{equation} 
which solves the PDE (\ref{gf-pde-backw}) with $q_1=y_1=0$\ie with
initial condition
\begin{equation} 
 \label{p-init} P(0,y) = \delta(y).
\end{equation} 
This function first appeared in the context of the
Sherrington-Kirkpatrick model in Ref.\ \cite{som81}.  Note that the
PDE for $P(q,y)$ is in fact a Fokker-Planck equation, producing a
nonnegative solution and conserving the norm $\int dy\, P(q,y)\equiv 1$
for all $q$-s.  This suggests the intuitive interpretation of $P(q,y)$
as probability density of $y$.

Now we are in the position to calculate the variation of the free
energy functional (\ref{fe-functional-spher-def-ext}).  As to the
energy term (\ref{fe-from-pde}), by varying the functions $f(q,y)$ and
$x(q)$ in the PDE (\ref{ppde1}) one obtains
(\ref{theta-with-source-pde}) with $\vartheta=\delta f$, $
\vartheta(1,y)=0$, and $h=\hf \beta (\ptl _yf)^2 \delta x$.  Hence
(\ref{theta-with-source-solve}) gives at $q=0, y=0$ the sought $\delta
f(0,0)/\delta x$.  The variation of $f_s[x(q)]$ can be calculated
straightforwardly, and, with the notation (\ref{p}), we arrive at
\begin{equation}
   F(q,[x(q)]) = \frac{2}{\beta} \frac{\delta f[x(q)]}{\delta x(q)} =
  \left( \int_0^q \frac{d\bar{q}}{\beta^2D(\bar{q})^2} - \alpha \int
  dy\, P(q,y)\,m(q,y)^2 \right)\label{var-fe-ext}.
\end{equation}

The stationarity condition in case $x(q)$ can be freely varied is thus
\begin{equation}
\label{stat1}
F(q,[x(q)]) = 0.
\end{equation}
If in an interval $I$ the $x(q)$ is supposed to have a plateau, we
differentiate by the plateau value of $x(q)$ to get
\begin{equation}
\label{stat2}
\int_I dx\, F(q,[x(q)]) = 0.
\end{equation}
In isolated points $q_r$ where plateaus meet (\ref{stat1}) should hold
pointwise. This summarizes the stationarity conditions for an
arbitrary order parameter function\ie arbitrary replica symmetry
broken scheme of Parisi type. 

We have seen in Sec.\ \ref{sec_storage} instances when the double,
thermal and quenched, average could be replaced by the replica average
(\ref{av}).  Replica averages can be calculated by the Green function
technique as described below.  The procedure can be viewed as the
generalization of the groundbreaking results from Refs.\
\cite{dl83,mv85}, where the local magnetization and some of its
moments in the Parisi phase of the Sherrington-Kirkpatrick model were
evaluated.

A simple case is when $\lav A(y_a)\rav$ is to be calculated for an
arbitrary function $A(y)$.  Because of the symmetry with respect to
the permutation of single replica indices we have $\lav A(y_a)\rav =
\lav n^{-1}\sum_{a=1}^n A(y_a)\rav$.  This quantity can be easily
evaluated if one replaces in (\ref{fe-spher1d}) $V(y)$ by
$V(y)+\lambda A(y)$, thus obtains $f_e({\sf Q};\lambda)$, and
calculates its initial slope at $\lambda=0$.  Reversing the limits
$n\to 0$ and $\lambda\to 0$ then using the first equality of
(\ref{phirrsb}) and Eq.\ (\ref{fe-from-pde}) we get
\begin{eqnarray}
\label{av-y2}
\lav A(y_a)\rav &=& \lim_{n\to 0} \left.\frac{1}{n}\,\frac{\ptl f_e({\sf
Q};\lambda)}{\ptl \lambda}\right|_{\lambda=0} = \left.\frac{\ptl
f(0,0;\lambda)}{\ptl \lambda}\right|_{\lambda=0} \nn \\ &=& \int dy\,
\frac{\delta f(0,0)}{\delta f(1,y)}\, A(y) = \int dy\, P(1,y)\, A(y).
\end{eqnarray} 
The third equality comes from the fact that $\lambda$ is in the
initial condition for $f(q,y)$ at $q=1$, and the last one comes from
the definitions for the Green function (\ref{gf-X}) and for $P(q,y)$
(\ref{p}).  Again, for the sake of brevity we gave the derivation for
continuous potential $V(y)$, however, the result holds also for
discontinuous ones.  Immediately follows from (\ref{repl-dist-stab})
the formula for the probability density of the local stabilities
\begin{equation}
\label{loc-stab-with-p}
\rho(y) = P(1,y),
\end{equation}
and thus from (\ref{ener-dist})
\begin{equation}
\label{e-with-p}
\varepsilon= \int dy\, P(1,y)\, V(y).
\end{equation}

From the practical viewpoint interesting are effective averages of
products like 
\begin{equation}
\label{av-y-s}
\lav A_1(y_{a_1})\,A_2(y_{a_2})\dots A_k(y_{a_k})\rav .
\end{equation}
One can show by using elementary properties of the Fourier
transformation that the average of a product of $x_a$-s over the
effective Boltzmann weight on the \rhs of (\ref{fe-spher1d}) can be
expressed as averages over functions of variables $y_a$-s.  Thus the
knowledge how to evaluate (\ref{av-y-s}) also resolves the problem of
averages of polynomials in $x_a$-s.  The latter quantities are of
importance because they appear when the replica free energy is
differentiated in terms of $q_{ab}$-s.

Here we shall only describe the recipe for calculating (\ref{av-y-s}),
details can be found in Ref.\ \cite{our-paper2}.  If $k=2$ then the
average depends on $q=q_{a_1 a_2}$ and is given by the formula
\begin{eqnarray}
C_{12}(q) &=& \lav A_1(y_{a_1})\,A_2(y_{a_2})\rav \nn \\ &=& \int
dy\,dy_2\,dy_3\, P(q,y)\, {\cal G}(q,y;1,y_1)\, {\cal G}(q,y;1,y_2)\,
A_1(y_1) \, A_2(y_2).
\label{av-yy}
\end{eqnarray}
Of such type is $\ptl f_e({\sf Q}) / \ptl q_{a_1 a_2}$ where
$A_1(y)=A_2(y)=i\cdot m(1,y)$, see (\ref{m}) for definition, whence we
obtain the second term in $F(q,[x(q)])$ given in
Eq. (\ref{var-fe-ext}).  This is related to the fact that the
stationarity condition can also be obtained by first differentiating
the replica free energy (\ref{fe-spher1b}) by $q_{ab}$ and then
equating the result to zero.  For $k=3$ suppose without restricting
generality that
\begin{equation}
\label{qq}
q=q_{a_1 a_3}=q_{a_2 a_3}<{\bar q}=q_{a_1 a_2}.
\end{equation}
Then we have
\begin{eqnarray}
\label{av-yyy}
C_{123}(q,{\bar q})&=&\lav
A_1(y_{a_1})\,A_2(y_{a_2})\,A_3(y_{a_3})\rav \nn \\ &=& \int
dy\,d{\bar y}\,dy_1\,dy_2\,dy_3\, P(q,y)\, {\cal G}(q,y;1,y_3)\, {\cal
G}(q,y;{\bar q},{\bar y})\, {\cal G}({\bar q},{\bar y};1,y_1)\,{\cal
G}({\bar q},{\bar y};1,y_2)\nn \\ && \times A_1(y_1) \, A_2(y_2) \,
A_3(y_3).
\end{eqnarray}
Special versions of the above formulas, for the case of the second and
third moments of the magnetization in the Sherrington-Kirkpatrick
model, were worked out in Refs.\ \cite{dl83,mv85}.  The integrals in
(\ref{av-yyy}) admit a simple graphic representation as shown on Fig.\
1.

\vskip0.5in
\begin{figure}[htb]
\begin{center}
\setlength{\unitlength}{3947sp}%
\begingroup\makeatletter\ifx\SetFigFont\undefined%
\gdef\SetFigFont#1#2#3#4#5{%
  \reset@font\fontsize{#1}{#2pt}%
  \fontfamily{#3}\fontseries{#4}\fontshape{#5}%
  \selectfont}%
\fi\endgroup%
\begin{picture}(3537,2235)(2989,-3811)
\thinlines
\special{ps: gsave 0 0 0 setrgbcolor}\put(4201,-3436){\line( 0,-1){150}}
\special{ps: grestore}\special{ps: gsave 0 0 0 setrgbcolor}\put(5026,-3436){\line( 0,-1){150}}
\special{ps: grestore}\special{ps: gsave 0 0 0 setrgbcolor}\put(4201,-2386){\line( 5, 2){1590.517}}
\special{ps: grestore}\special{ps: gsave 0 0 0 setrgbcolor}\put(5101,-2011){\line( 3,-1){675}}
\special{ps: grestore}\put(4201,-3286){\makebox(0,0)[b]{\smash{\SetFigFont{12}{14.4}{\rmdefault}{\mddefault}{\updefault}\special{ps: gsave 0 0 0 setrgbcolor}$q$\special{ps: grestore}}}}
\put(6226,-1786){\makebox(0,0)[b]{\smash{\SetFigFont{12}{14.4}{\rmdefault}{\mddefault}{\updefault}\special{ps: gsave 0 0 0 setrgbcolor}$A_1$\special{ps: grestore}}}}
\put(6226,-2311){\makebox(0,0)[b]{\smash{\SetFigFont{12}{14.4}{\rmdefault}{\mddefault}{\updefault}\special{ps: gsave 0 0 0 setrgbcolor}$A_2$\special{ps: grestore}}}}
\put(6226,-3136){\makebox(0,0)[b]{\smash{\SetFigFont{12}{14.4}{\rmdefault}{\mddefault}{\updefault}\special{ps: gsave 0 0 0 setrgbcolor}$A_3$\special{ps: grestore}}}}
\put(4501,-2086){\makebox(0,0)[b]{\smash{\SetFigFont{12}{14.4}{\rmdefault}{\mddefault}{\updefault}\special{ps: gsave 0 0 0 setrgbcolor}$\cal G$\special{ps: grestore}}}}
\put(5026,-3286){\makebox(0,0)[b]{\smash{\SetFigFont{12}{14.4}{\rmdefault}{\mddefault}{\updefault}\special{ps: gsave 0 0 0 setrgbcolor}$\bar q$\special{ps: grestore}}}}
\special{ps: gsave 0 0 0 setrgbcolor}\put(4201,-2386){\line( 5,-2){1590.517}}
\special{ps: grestore}\special{ps: gsave 0 0 0 setrgbcolor}\put(3001,-2386){\line( 1, 0){1200}}
\special{ps: grestore}\special{ps: gsave 0 0 0 setrgbcolor}\put(3001,-3511){\line( 1, 0){2775}}
\special{ps: grestore}\special{ps: gsave 0 0 0 setrgbcolor}\put(3001,-3436){\line( 0,-1){ 75}}
\special{ps: grestore}\special{ps: gsave 0 0 0 setrgbcolor}\put(3001,-3436){\line( 0,-1){150}}
\special{ps: grestore}\special{ps: gsave 0 0 0 setrgbcolor}\put(5776,-3436){\line( 0,-1){150}}
\special{ps: grestore}\special{ps: gsave 0 0 0 setrgbcolor}\put(5776,-3511){\line( 1, 0){300}}
\special{ps: grestore}\special{ps: gsave 0 0 0 setrgbcolor}\put(6076,-3511){\line(-2, 1){150}}
\special{ps: grestore}\special{ps: gsave 0 0 0 setrgbcolor}\put(6076,-3511){\line(-2,-1){150}}
\special{ps: grestore}\put(5776,-3811){\makebox(0,0)[lb]{\smash{\SetFigFont{10}{12.0}{\rmdefault}{\mddefault}{\updefault}\special{ps: gsave 0 0 0 setrgbcolor}1\special{ps: grestore}}}}
\put(3001,-3811){\makebox(0,0)[lb]{\smash{\SetFigFont{10}{12.0}{\rmdefault}{\mddefault}{\updefault}\special{ps: gsave 0 0 0 setrgbcolor}0\special{ps: grestore}}}}
\end{picture}
\end{center}
\caption{Correlation function $C_{123}(q,{\bar q})$.  The right
endpoints correspond to the functions to be averaged and lines are
Green functions (only one of them is labeled by $\cal G$ in the
figure).  All nodes, including the right endpoints, have $y$-s which
should be integrated over.  The leftmost line is the function
$P(q,y)$, or equivalently, this line is also a Green function and a
$\delta(y)$ is associated with the leftmost endpoint.}
\label{c-x2}
\end{figure}
\vskip0.5in

The cases considered above are the effective average for $k=1$, given
by Eq.\ (\ref{av-y2}), represented by one line, and for $k=2$, as
calculated in (\ref{av-yy}), represented by a fork with one handle and
two branches.  Averages of more than three functions can be
analogously constructed, for a given $k$ a graph has $k+1$ ``legs''.
Obviously there are two topologically possible graphs for $k=4$,
depending on the overlaps $q_{a_{i} a_{j}}$, and more for larger
$k$-s.

The ability to calculate $k=4$ effective averages allows us to study
linear stability of the replica free energy (\ref{fe-spher1b}) at the
stationary $\sf Q$.  Using the results of Ref.\ \cite{tdk94} on
ultrametric matrices we expressed the so called replicon eigenvalues
in terms of Green functions.  While a general proof of the fact that
there are no negative eigenvalues in the case of continuous replica
symmetry breaking\ie when $x(q)$ has a continuously increasing
segment, is not available, in the high temperature limit
\cite{our-paper,our-paper2} we confirmed the absence of linear
instability against replicons whenever we encountered such a
stationary state.  For any temperatures we recovered analytically the
zero eigenvalues, corresponding to Goldstone modes, as well as the
lowest order Ward-Takahashi identities predicted by algebra
\cite{dtk98}.

The generalization of (\ref{av-y-s}) to non-factorizable functions is
straightforward.  For example, such functions would simply replace the
products of $A_k$-s in (\ref{av-yy}) and (\ref{av-yyy}).

\section{Low temperature results}
\label{lt} 

In Ref.\ \cite{our-paper} we have shown that in the high temperature
limit\ie for $\alpha,\beta\to\infty$ with $\gamma=\alpha \beta^2$
finite, the problem simplifies to the extent that if $x(q)$ has a
continuously increasing segment, this can be given in a closed
analytic form.  In that limit the problem becomes equivalent to the
spherical, multi-$p$-spin interaction spin glass \cite{nie95}, where a
similar observation has been made.  Four different phases have been
found \cite{our-paper} for the error counting potential (\ref{gen-v}):
for small $\gamma$ replica symmetry holds, and for $|\kappa|<2$ and
large $\gamma$ there is a Parisi phase with a single continuously
increasing segment of $x(q)$ between the trivial plateaus $x\equiv0$
and $x\equiv1$.  When $|\kappa|>2$ there is also a narrow one-step
replica symmetry broken regime, and for large enough $\gamma$-s
equilibrium is characterized by an $x(q)$ that is a concatenation of a
nontrivial plateau and a continuously increasing segment.  While for
$T>0$ there cannot be error free storage, it is plausible to conceive
the replica symmetric regime as the continuation of the $T=0$ phase of
perfect storage and the symmetry broken phases as the analog of the
regime at $T=0$ beyond capacity.

In the case of a $V(y)$ potential that vanishes for $y>\kappa$ the
limit of capacity is given for $\kappa\geq 0$ at $T=0$ by \cite{eg88}
\begin{equation}
\label{c}
\alpha _c(\kappa)= \left(\int_{-\infty}^{\kappa} Dt\,(\kappa -t)^2
\right)^{-1},
\end{equation}
as it follows from the replica symmetric solution when $q\to 1$.  This
formula also gives the limit of the de Almeida-Thouless (AT) local
stability \cite{bou94} in the case of the potential (\ref{gen-v}).
For $\kappa =0$ one has $\alpha _c=2$, and, for increasing $\kappa$,
$\alpha _c(\kappa)$ understandably decreases.  As it has been already
mentioned, beyond capacity none of the finite-step replica symmetry
breaking schemes gives a locally stable equilibrium state \cite{ws96}.
Thus in this regime the order parameter function is no longer of the
step-like form of (\ref{xq}), rather it has a continuously increasing
part.  This makes it necessary to numerically solve the extremization
problem (\ref{fe-spher-var-ext}).

The ground state (T=0) has its special scaling properties. The PDE
(\ref{ppde1}) stays meaningful if for $q<1$ the function $\beta x(q)$
does not diverge, implying that $x(q)$ goes to zero.  Given the
meaning of $x(q)$ as the probability of a $q$ being in the interval
$(0,q)$ \cite{sgrev87}, we can say that at $T=0$ the overlap $q$ is
$1$ with probability 1 for all $\alpha>\alpha_c$.  Nevertheless, a
physically meaningful order parameter is obtained after scaling by
$\beta$.  Firstly we introduce for $T>0$ the parameters of the classic
Parisi shape as
\begin{equation}
\label{xq-parisi-phase-def}
    \left. \begin{array} {ll} x(q) \equiv 0 & \text{~~if~} 0\leq q
	<q_{(0)} \\ {\dot x}(q)>0 \text{~and~} x_{(0)}< x(q) < x_{(1)}
	& \text{~~if~} q_{(0)}< q < q_{(1)} \\ x(q) \equiv 1 &
	\text{~~if~} q_{(1)}< q \leq 1 . \end{array} \right.
\end{equation}
The scaled quantities 
\bml
\label{scaling}
\begin{eqnarray}
\label{scaling1}
q(t) &=& q_{(1)} - \left( q_{(1)} - q_{(0)}\right) \left( 1 - (1 +
q_{(1)})\,t+ q_{(1)}t^2\right), ~~~ 0\le t \le 1 , \\
\label{scaling2}
\xi(t) &=& \beta x(q(t))\,\dot{q}(t), \\
\label{scaling3}
\eta&=& \beta \left(1-q_{(1)}\right), \\
\label{scaling4}
\Delta(t) &=& \beta D(q(t)) = \int_t^1 \xi(\bar{t}) d\bar{t} + \eta, 
\end{eqnarray}
\eml are expectedly regular even in the $T\to 0$ limit, when
$q_{(1)}\to 1$.  Note that there is an arbitrariness in the
parametrization by $t$, the main features being that $q(0)=q_{(0)}$,
$q(0)=q_{(0)}$, $q(1)=q_{(1)}$, and ${\dot q}(1)=0$.  With this
parametrization the PDE (\ref{ppde}) becomes \bml
\label{ppde-rescaled}
\begin{eqnarray}
 \ptl _t f(t,y) &= & -\hf \,\dot{q}(t)\, \ptl _y ^2 f(t,y) + \hf\,
   \xi(t)\, \left(\ptl _y f(t,y) \right)^2 , \label{ppde-rescaled1} \\
   f(1,y) &= &- \frac{1}{\beta} \ln \int Dz\, e^{-\beta
   V\left(y+z\sqrt{1-q_{(1)}} \right)}.
   \label{ppde-rescaled2}
\end{eqnarray}
\eml
For $T=0$ the initial condition becomes 
\begin{equation}
 \label{ppde-rescaled2-T0} \left. f(t=1,y)\right|_{T=0}=\min_{\bar{y}}
  \left(V(\bar{y}) + \frac{(y-\bar{y})^2}{2\eta} \right) = \left\{
  \begin{array} {ll} 1 & \text{if~} y \le \kappa-\sqrt{2\eta} \\
  \frac{(\kappa -y)^2}{2\eta} & \text{if~} \kappa-\sqrt{2\eta} \le y
  \le \kappa \\ 0 & \text{if~} y \ge \kappa , \end{array} \right.
\end{equation}
where (\ref{gen-v}) was substituted to get the second equality.  By
Gaussian integration hence the replica symmetric solution can be
obtained \cite{eg88,gd88,gg91}.  Note that $f(t=1,y)$ is a continuous
function even though $V(y)$ is the step function.  Although we used
the same symbols for functions of $q$ and $t$, misunderstanding are
avoided by our marking which argument we mean.

The PDE for $P(q,y)$ follows from the definition (\ref{p}) and from
the evolution equation of the Green function (\ref{gf-pde-backw}).
The latter can be properly rescaled according to (\ref{scaling}),
yielding in principle the solution $P(t,y)$.  The probability density
of local stabilities is obtained by evolving $P(t=1,y)=P(q=q_{(1)},y)$
to $P(q=1,y)$.  At $T=0$ with (\ref{gen-v}) we have
\begin{equation}
\label{local-stability-distr-T=0-theta-v}
    \rho(\Delta) = \left\{ \begin{array} {ll} P(1, \Delta) &
	\text{if~} \Delta \leq \kappa-\sqrt{2\eta} \\ 0 & \text{if~}
	\kappa-\sqrt{2\eta} < \Delta < \kappa \\ P(1, \Delta) +
	\delta( \Delta-\kappa)\,\int_{\kappa-\sqrt{2\eta}}^{\kappa}
	d\bar{y}\, P(1,\bar{y})\, & \text{if~} \Delta \ge \kappa
	, \end{array} \right.
\end{equation}
where $P(1, \Delta)=P(t=1, \Delta)$ is understood.  For arbitrarily
small $T>0$ the gap in the support of $\rho( \Delta)$ immediately
vanishes.  For details we again refer to \cite{our-paper2}.

The numerical extremization was done with complementing the free
energy functional (\ref{fe-functional-spher-def-ext}) by constraints.
The PDE (\ref{ppde-rescaled}) was added giving rise to a Lagrange
multiplier field.  This field can be shown to be just $P(t,y)$, see
Refs.\ \cite{our-paper,our-paper2,sd84}.  Further technical
requirements are $x(q_{(0)})\geq 0$, $x(q_{(1)})\leq 1$, and ${\dot
x}(q)\geq 0$ for $q_{(0)}<q<q_{(1)}$, which were taken into account by
soft constraints.  A few results for the Gardner-Derrida potential
(\ref{gen-v}), with $\kappa=0$, $\alpha=3$, are displayed on Figs.\ 2
and 3 for various low temperatures.  Note that the point
$(\kappa=0,\alpha=3)$ lies beyond capacity, while beyond about
$\beta^{-1}=T=0.2$ the RS solution satisfies the AT
stability condition \cite{our-paper2}.
  
On Fig. 2 the scaled order parameter function $\beta x(q)$ is shown.
For small temperatures the replica symmetric solution is AT unstable,
and we indeed obtain the Parisi form (\ref{xq-parisi-phase-def}) for
$x(q)$.  At $q_{(0)}$ near $0.75$ the functions $x(q)$ jump to zero
and remains there as $q$ further decreases.  The upper plateau with
$x(q)\equiv 1$ starts at $q_{(1)}$.  The consistency of the scaling
(\ref{scaling}) is confirmed by our finding at $T=0$ finite values for
$\beta x(q)$ if $q<1$ and for $\eta\approx 0.26$.  Interestingly, the
curved segment of $\beta x(q)$ does not change much with increasing
temperature, the main effect being the decrease of $q_{(1)}$.

\vskip0.1in \centerline{\psfig{figure=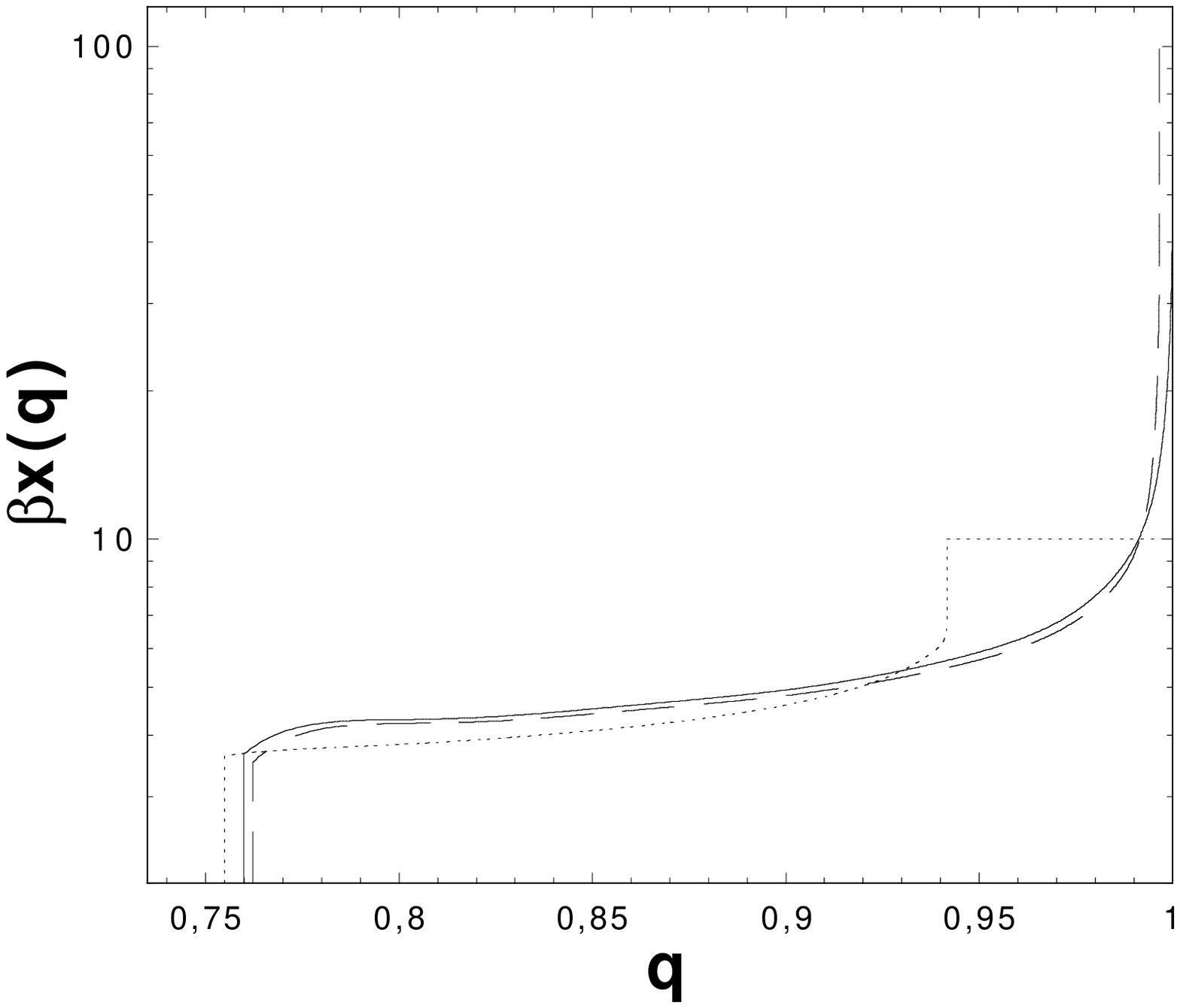,
 width=3in,height=3in,angle=0}} \vskip0.2in {\small FIG. 2. Scaled
 order parameter function $x(q)$ for $\kappa=0$, $\alpha=3$ at $T=0$
 (solid), $T=0.01$ (dashed), and $T=0.1$ (dotted).}  \vskip0.3in


The local stability distribution $\rho(\Delta)$ of
(\ref{repl-dist-stab}) is displayed on Fig.\ 3.  It was numerically
obtained for $T=0$ from (\ref{local-stability-distr-T=0-theta-v}) and
for $T>0$ from the original formula (\ref{loc-stab-with-p}), with the
same parameter values as in Fig.\ 3.  For the sake of better
visibility of the other details, the very high peaks near
$\Delta=\kappa=0$ for $T=0.1$ and $T=0.01$ as well as the
corresponding $\delta$-peak for $T=0$ have been omitted from this
plot.  For $|\Delta|>3$ the curves approach zero very quickly.  A true
gap with $\rho (\Delta) = 0$ to the left of $\Delta=0$ develops only
if $T=0$, while for the positive $T$-values $\rho(\Delta)$ is positive
albeit small there.  While the gap at $T=0$ is present in $R=0,1,2$
step replica symmetry breaking schemes, see Refs.\
\cite{gg91,mez93,ws96} respectively, in all these cases a jump appears
near the lower edge.  This can be associated with the thermodynamic
instability of those saddle points \cite{eng99}.  Our present solution
gives linearly vanishing $\rho(\Delta)$ at the lower edge, signaling
the absence of replicon instability \cite{our-paper2}.

\vskip0.1in \centerline{\psfig{figure=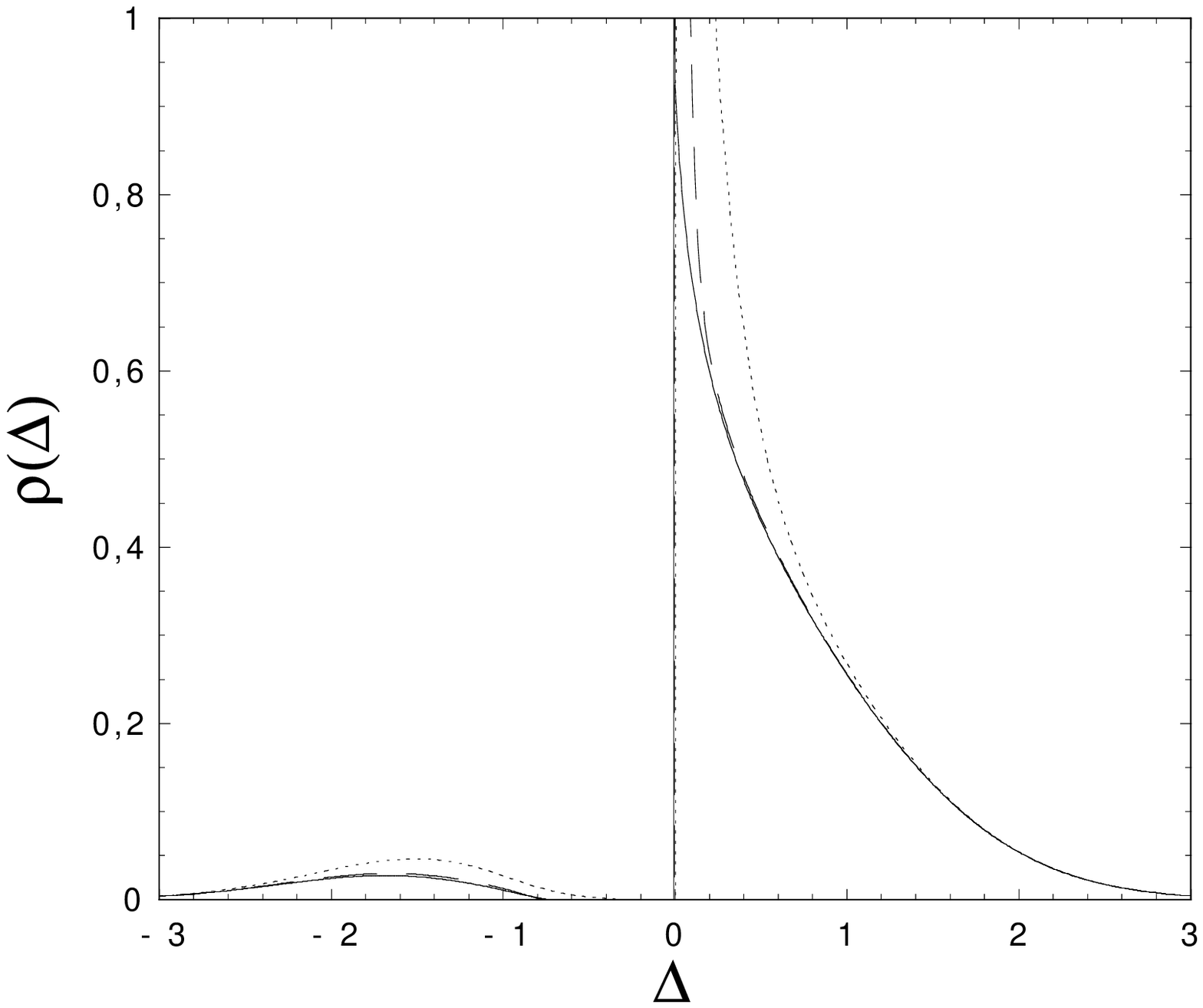,
width=4in,height=3in,angle=0}} \vskip0.2in {\small FIG. 3.  Density of
local stabilities $\rho(\Delta)$ from theory for $\kappa=0$,
$\alpha=3$ at $T=0$ (solid), $T=0.01$ (dashed), and $T=0.1$ (dotted).}
\vskip0.3in

In order to compare theory with practice we performed a medium scale
simulation.  The standard Hebbian algorithm was modified by Wendemuth
\cite{wen95a} to provide convergence for negative stabilities.  Since
we chose $\kappa=1$, the final steps during stabilization of a pattern
went on with $\Delta>0$, so in our case the modification was not
essential.  The algorithm goes as follows.  Firstly random patterns
(\ref{patterns}) are generated uniformly from an interval centered
about zero and normalized as $\sum_{k=1}^N (S^{\mu}_k)^2=N$, and all
outputs $\xi^{\mu}$ are taken uniformly $1$.  This does not restrict
generality since $S^{\mu}_k$ have random signs.  The initial coupling
vector is set to be proportional to
\begin{equation}
\label{initj}
J_k(0)\propto \sum_{\mu=1}^M S_k^{\mu},
\end{equation}
such that its Eucledian norm is $N$.  In the $t$-th step of the
algorithm one calculates the local stabilities
\begin{equation}
\label{locstab}
\Delta^{\mu}(t) = \frac{{\bf J}(t)\cdot {\bf S}^{\mu}}{|{\bf J}(t)|}
\end{equation}
and selects the least unstable pattern\ie the one with the largest
$\Delta^{\mu}(t)<\kappa$.  Let us denote its index by $\mu_0$, whose
argument $t$ we omit.  Next one augments the couplings as follows.  If
$\Delta^{\mu_{0(t)}}(t)>0$ then
\begin{equation}
\label{update1}
{\bf J}(t+1) = {\bf J}(t) + \lambda {\bf S}^{\mu_{0}},
\end{equation}
and if $\Delta^{\mu_{0}(t)}(t)<0$ we have following Wendemuth
\begin{equation}
\label{update2}
{\bf J}(t+1) = {\bf J}(t) + \lambda \left( {\bf S}^{\mu_{0}} +{\bf
J}(t) \frac{N/|{\bf J}(t)| - \Delta^{\mu_{0}}(t)}{|{\bf J}(t)| -
\Delta^{\mu_{0}}(t)} \right).
\end{equation}
Here $\lambda$ is the gain parameter, the overall scale of increments
of $\bf J$.  In Ref. \cite{wen95a} the gain parameter was
$\lambda=N^{-3/2}$, after experimentation we chose $\lambda=N^{-1}$.
Such an increase in the gain parameter did not endanger, rather sped
up convergence.  At time $t+1$ we again look for the least unstable
pattern, and so on.  The algorithm goes on until it gets stuck with
one pattern that we are not able to stabilize in a reasonable time.
The intuitive idea behind the algorithm is that since an unstable
pattern counts as error irrespective of the distance of the stability
parameter $\Delta^\mu$ from $\kappa$, one assumes that it is the
easiest to stabilize the pattern with $\Delta^\mu$ closest to
$\kappa$.  So one may hope that thus the largest possible number of
patterns can be stabilized.

The program ran on $28$ PC-s in parallel, each having an AMD K6
processor of $333$ MHz, during about one day.  Fig.\ 4 shows both the
theoretical curve and the results of the simulation for
$\alpha=\kappa=1$, a point known to fall beyond the capacity curve
(\ref{c}).  The full line is the result of numerical extremization of
the free energy functional in the way Fig.\ 3 was obtained.  The Dirac
delta peak of the theoretical probability density at $\kappa$ is not
illustrated.  The discontinuous lines represent the histograms for the
local stabilities from simulation for two sizes, $M=N=500$ and $1000$,
after normalization.

\vskip0.1in \centerline{\psfig{figure=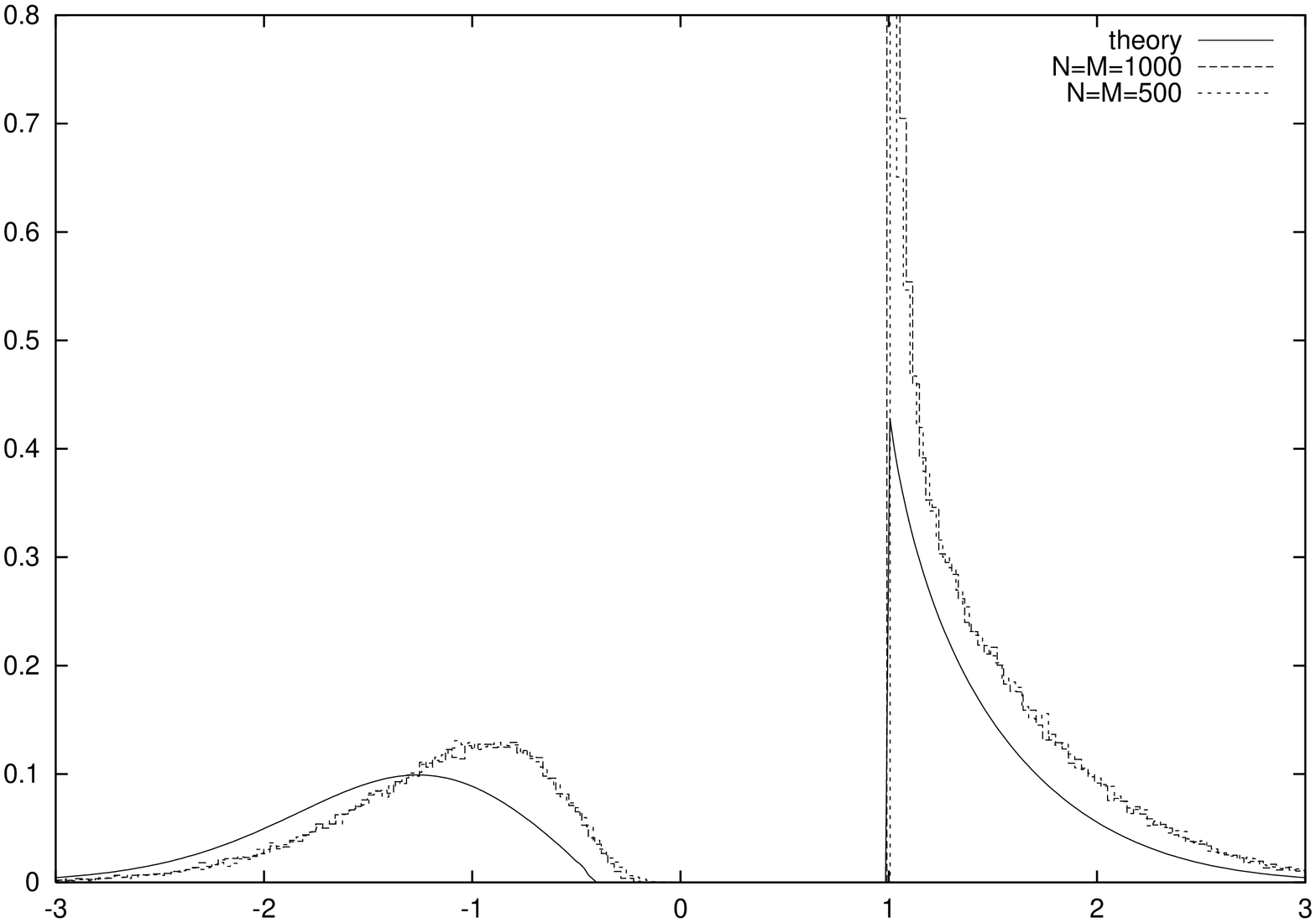,
width=3in,height=4in,angle=270}}\vskip0.1in {\small FIG. 4. Density of
local stabilities $\rho(\Delta)$ at $\alpha=\kappa=1$, axes as in
Fig.\ 3.  The theoretical prediction is given by the full line.  The
two empirical densities are normalized histograms, taken with
$M=N=500$ and $1000$.}  \vskip0.3in
 
The closeness of the two histograms demonstrates that size effects
were probably not the cause for the systematic difference between
theory and the numerical experiment.  A possible ground for the
discrepancy is that the algorithm may have been halted prematurely.
However, the time necessary for the stabilization of patterns was
allowed to grow for each subsequent pattern, and the algorithm was
ended only when stabilization did not occur even within the multiple
of such an extrapolated time.  Another possible reason for the
deviation may be that the algorithm got stuck in a ``local optimum''
without being able to globally maximize the number of stable patterns.
In this regard several modified initial conditions were tested but the
number of stabilized patterns did not grow in the end.  A source of
concern can be that the built in random number generator of the C
compiler was used; we did not test other routines for this purpose.
As to the algorithm, despite its intuitive appeal, there is no proof
that it would be able to globally minimize the Hamiltonian
(\ref{hamiltonian}) with error measure (\ref{gen-v}).  Furthermore, it
is likely that with the present parameters the learning task is an
NP-complete problem \cite{hkp91,wen95a,wen95b}, thus explaining
imperfect convergence.

We emphasize that the results represent a significant improvement with
respect to the earlier simulation in Ref. \cite{wen95b}.  The error
per example $\varepsilon$ found in \cite{wen95b} is about $0.21$,
while the present data correspond to $0.15$ and theory predicts
$0.1358$.  Thus the deviation between simulation and theory has been
decreased by 80\%.  That means that we stabilized more patterns than
\cite{wen95b}, although, given the difference between the theoretical
and simulation results, we still could not find the global optimum.
Furthermore, an important feature of the density $\rho(\Delta)$ is
that it should continuously vanish, with a nonzero slope, at the lower
edge of the gap.  This property is reproduced by the simulation data,
in a sharper fashion with the larger $M=N=1000$ size, while the value
of the edge remains slightly overestimated.

\acknowledgments
 
The authors acknowledge support by OTKA grant No.\ T017272 (G. Gy.)
and from a special grant for young scientists from the University of
Augsburg and by the State of Bavaria within the postgraduate scheme
Graduiertenkolleg GRK283 ``Nonlinear Problems in Analysis, Geometry,
and Physics'' (P. R.).  It is a pleasure to thank F. Csikor and
Z. Fodor for offering us for the simulation their PC farm, supported
by grants Nos. OTKA-T22929 and FKFP-0128/1997. Thanks are due to
F. P\'azm\'andi for his pointing out and discussing with us the
problem of a discontinuity in the potential.

\end{document}